\documentclass{ab2018}
\usepackage{url}
\usepackage{graphicx}
\usepackage{natbib}

\newcommand{\SII}{[S\,II]}
\newcommand{\OIII}{[O\,III]}
\newcommand{\NII}{[N\,II]}
\newcommand{\HII}{H\,II}
\newcommand{\HI}{H\,I}

\newcommand{\Ha}{H$\alpha$}
\newcommand{\Hb}{H$\beta$}
\newcommand{\kms}{\,\mbox{km}\,\mbox{s}^{-1}}

\newcommand{\NIIHa}{$I$([N\,II])/$I$(H$\alpha$)}
\newcommand{\OIIIHb}{$I$([O\,III]\,$\lambda$\,5007)/$I$(H$\beta$)}

\begin{document} 

\title{Ionized gas in the NGC\,3077 galaxy}

\author{D.~V. Oparin$^1$,  {O.~V.}~Egorov$^{2,1}$ and A.V.~Moiseev$^{1,2}$}

\institute{
$^1$ Special Astrophysical Observatory, Russian Academy of Sciences, Nizhnij Arkhyz, 369167  Russia\\
$^2$ Sternberg Astronomical Institute, M.V.Lomonosov Moscow State University, Moscow, 119234 Russia
}

\titlerunning{Ionized gas in the NGC\,3077}

\authorrunning{Oparin et al.}

\date{July 6, 2020/Revised: July 28, 2020/Accepted: July 28, 2020}
\offprints{Dmitry Oparin  \email{doparin2@gmail.com} }

\abstract{
The nearby dwarf galaxy NGC\,3077 is known for its peculiar
morphology, which includes numerous dust lanes and emission-line
regions. The interstellar medium in this galaxy is subject to
several perturbing factors. These are primarily the central
starburst and tidal structures in the M\,81 group. We present a
comprehensive study of the state of ionization, kinematics, and
chemical composition of ionized gas in NGC\,3077, including both
star-forming regions and diffuse ionized gas (DIG) at the
periphery. We study gas motions in the \Ha{} line via
high-resolution ($R\approx15\,000$) 3D spectroscopy with
the scanning Fabry-Perot interferometer installed into SCORPIO-2
instrument attached to the 6-m telescope of the Special
Astrophysical Observatory of the Russian Academy of Sciences.
Images in the main optical emission lines were acquired with
MaNGaL photometer with a tunable filter at the 2.5-m
telescope of the Caucasian Mountain Observatory of Sternberg
Astronomical Institute of M.V. Lomonosov Moscow State University.
We also used SCORPIO-2 to perform long-slit spectroscopy of the
galaxy with a resolution of  $R\approx1\,000$. Our estimate of the
gas metallicity, $Z=0.6Z_\odot$, is significantly lower than the
earlier determination, but agrees with the
``luminosity--metallicity'' relation. Spatially resolved
diagnostic diagrams of the emission-line ratios do not show
correlations between the gas ionization state and its velocity
dispersion, and this is most likely due to strong ionization by
young stars, whereas the contribution of shocks to the excitation
of emission lines is less important.  We also studied the
locations of multicomponent  \Ha{} profiles and provide arguments
suggesting that they are mostly associated with individual
kinematic components along the line of sight and not with
expanding shells as it was believed earlier. We also observe there
a combination of wind outflow from star-forming regions and
accretion from interstellar gas clouds in the M\,81 group.
}

\maketitle
\section{INTRODUCTION}
\label{sec:intro}

\begin{figure*}[!]
\centerline{
\includegraphics[height=0.6\linewidth]{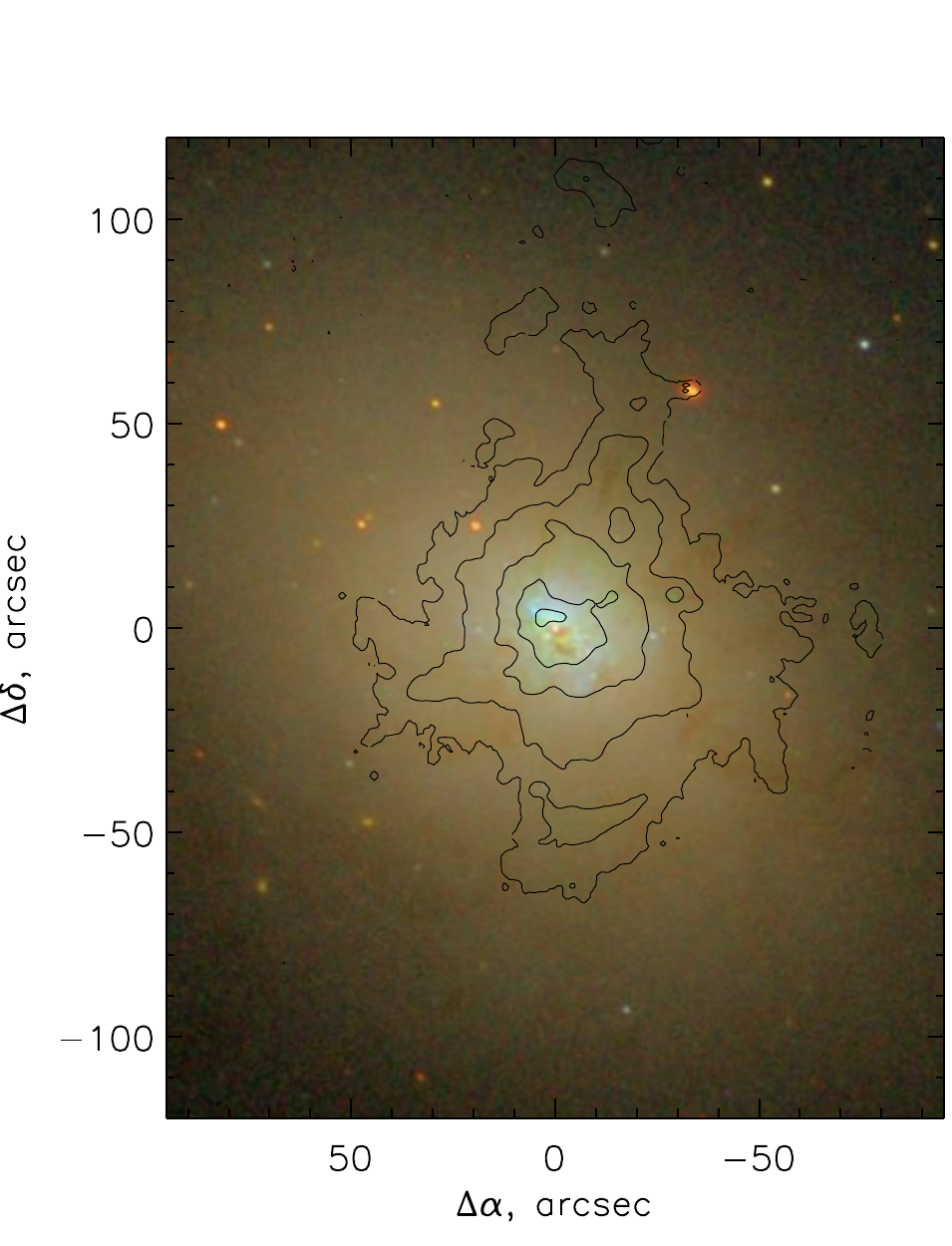}
\includegraphics[height=0.6\linewidth]{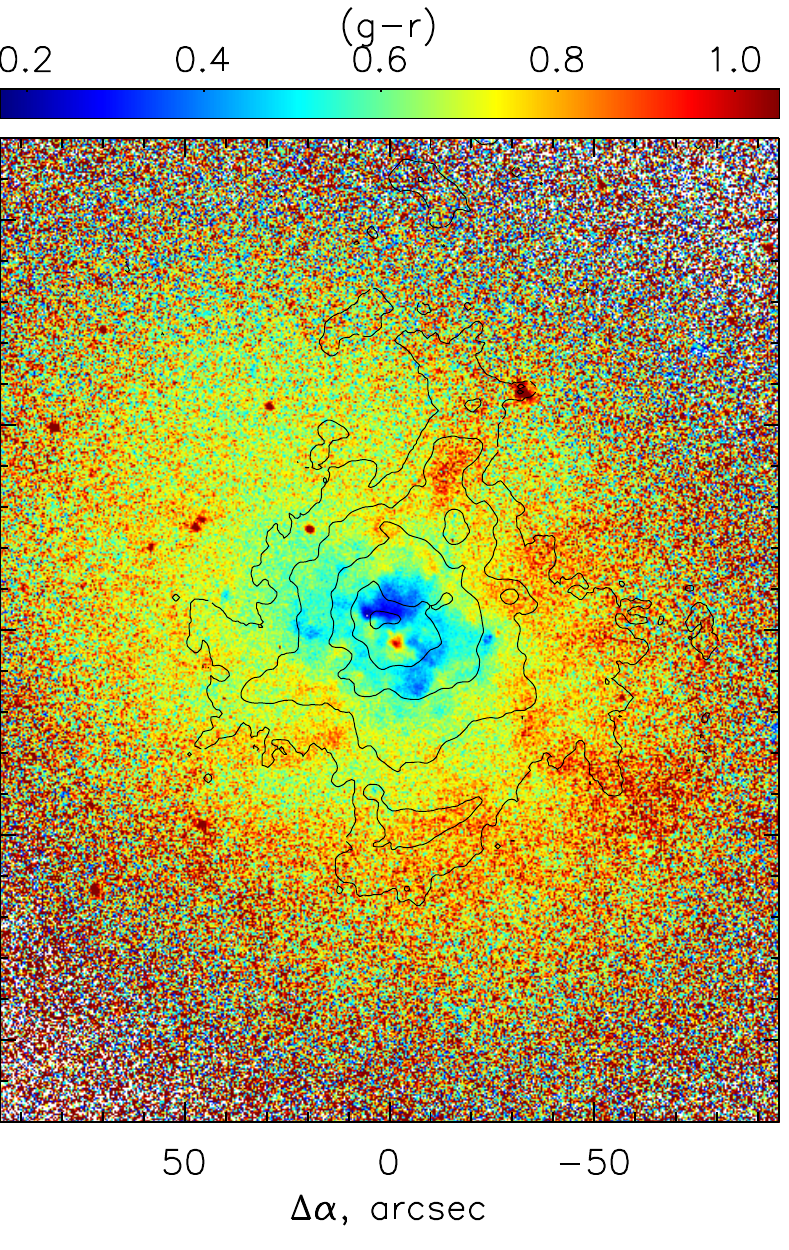}
}
\centerline{
\includegraphics[height=0.6\linewidth]{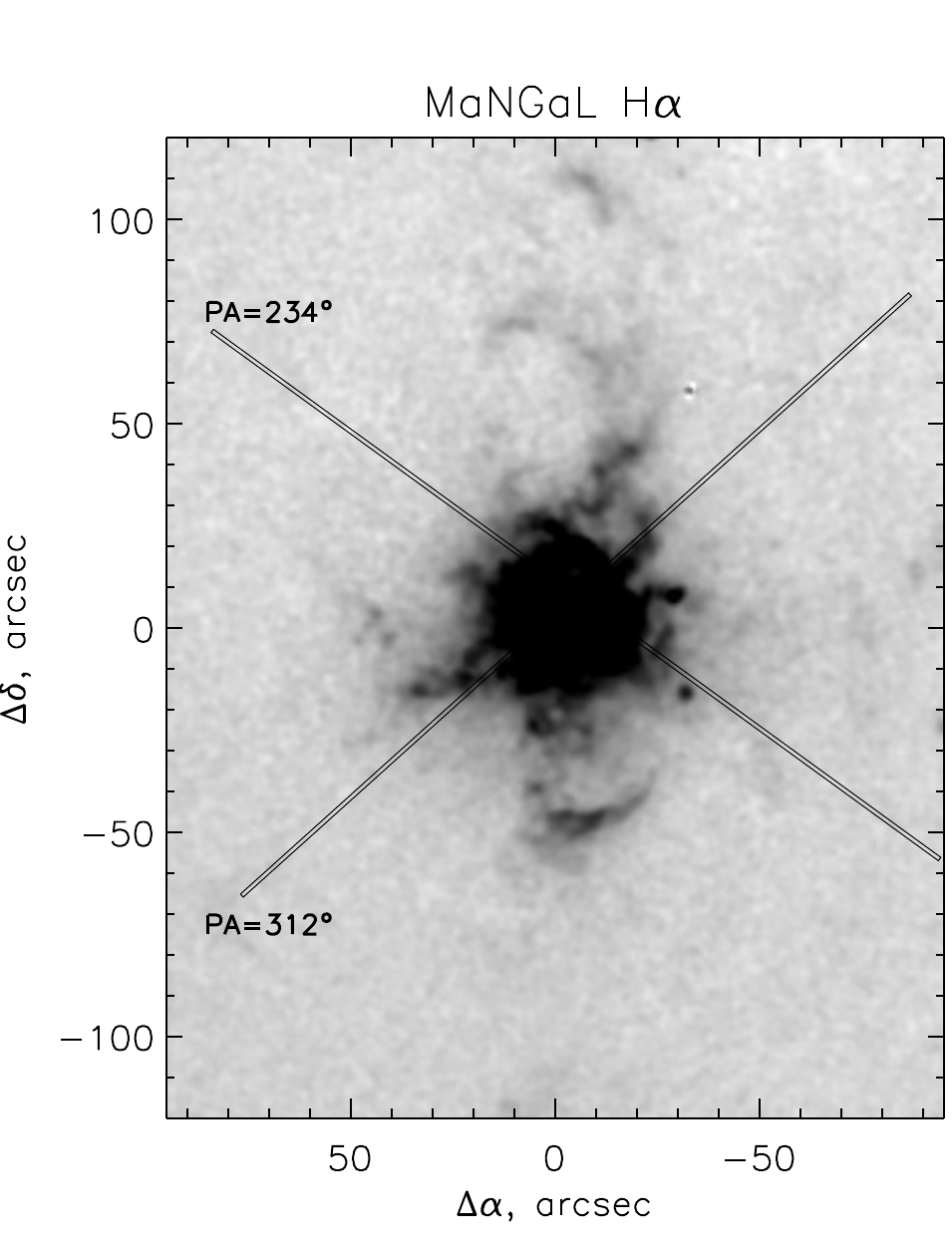}
\includegraphics[height=0.6\linewidth]{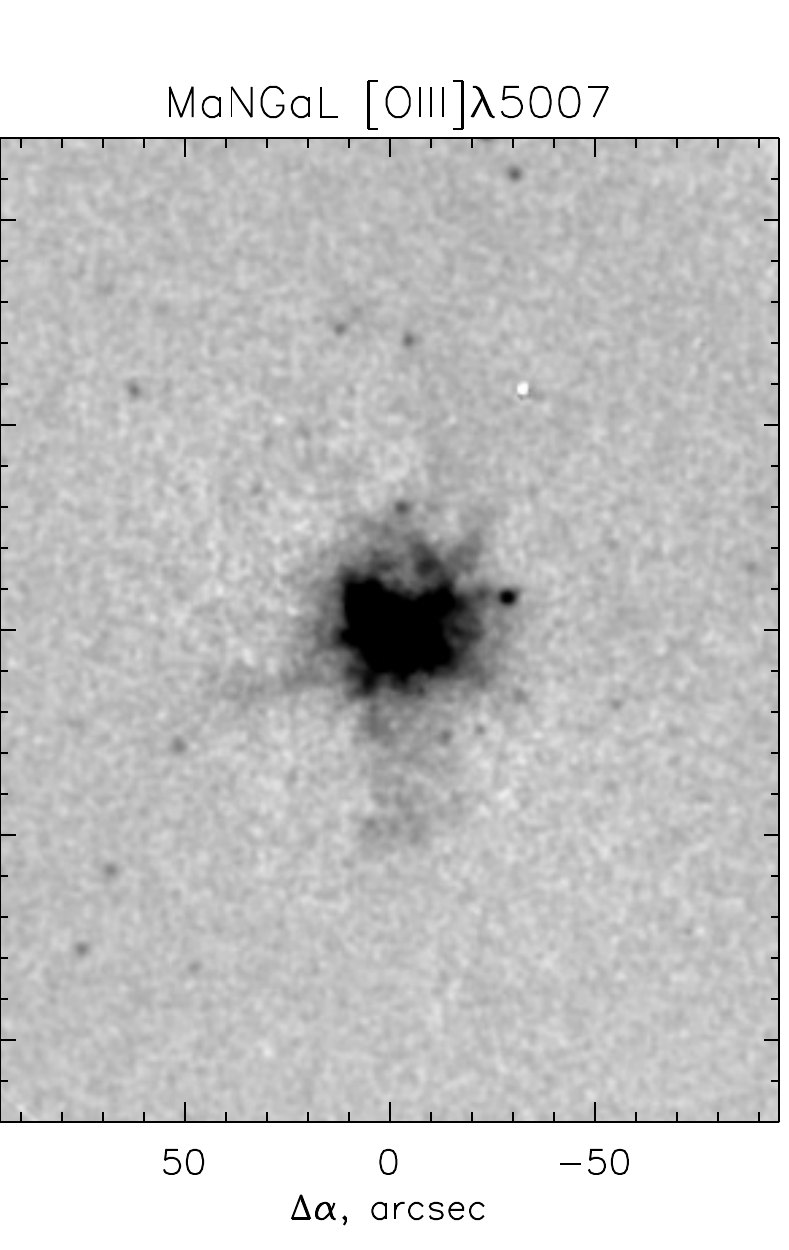}
}
\caption{The top row: composite color image of NGC\,3077
adopted from SDSS DR15 sky survey (left) and $(g-r)$ color index
map (right). Also shown are the  H$\alpha$-line contours based on the
results of our FPI observations made with the 6-m telescope. The
bottom rows:  images acquired with MaNGaL on the 2.5-m
telescope in the \Ha{} emission line with the positions of the  SCORPIO-2 slits (left) and   in the \OIII{} emission line (right).}
    \label{fig:sdss3077}
\end{figure*}

NGC\,3077 is a dwarf galaxy, which is a member of the M\,81 group.
Its absolute magnitude and adopted distance are  $M_B=-17.62 $ and
3.85~Mpc, respectively \citep{Kaisina2012}. NGC\,3077 stands out
among nearby galaxies by its peculiar morphology: numerous
ionized-gas filaments and the associated dust lanes, which show up
conspicuously in optical images (Fig.~\ref{fig:sdss3077}).  HI
line observations show that the galaxy is located near a giant
tidal structure, which connects it to  M\,81 and
M\,82~\citep{Yun1994Natur.372..530Y,Sorgho2019MNRAS.486..504S},
and the entire system is sometimes called the  ``M\,81
triplet''~\citep{deBlok2018ApJ...865...26D}.  On the southern side
of the galaxy a chain of clusters of blue stars and compact \HII{}
regions can be seen---it is the ``Garland'' tidal
galaxy~\citep{Karachentsev1985,Makarova2002}, which is located in
a region of high \HI{} density. The rate of ongoing star formation
in NGC\,3077 estimated from \Ha, luminosity and the HI mass are
equal to \mbox{$SFR\approx0.1\,M_{\odot}$/yr} and
\mbox{$M_{\rm{H\,I}}=6.3\times10^8\,M_{\odot}$},
respectively~\citep{Karachentsev2007, Karachentsev2013}, which is
quite significant for a dwarf galaxy.

\citet{Martin_1998} used  \Ha-echellegrams to demonstrate that
NGC\,3077 contains regions of broadening and bifurcation of
emission lines interpreted as manifestations of several expanding
shells of ionized gas. An estimate of virial velocities showed
that their energy is insufficient for ejecting matter from the
galaxy. The rotation curve of the galaxy was determined based on
\HI{} data and the rotation velocity was found to amount to
$50\kms$. At the same time,  \Ha{} observations show no signs of
rotation~\citep{GHASPVI}.

\citet{Ott2005} demonstrated the presence of a cluster of compact
X-ray sources at the center of the galaxy, which are spatially
associated with one of the expanding shells earlier discovered
optically. It can be suggested that we are dealing with supernova
remnants. The above authors point out that shells observed in
emission lines are filled with optically thick gas. Note that
unlike other seven dwarf starburst galaxies discussed in the above
paper NGC\,3077 has close-to-solar
metallicity~\citep{Storchi-Bergmann1994} and rather low X-ray
brightness.

The diagrams of the flux ratios of bright optical emission lines
(\OIII/\Hb, \NII/\Ha{}, \SII/\Ha)---the so-called BPT diagrams
proposed in the classical paper by \citet*{1981BPT} and later further developed
in \citet{Veilleux1987} are an important tool for studying the gas
ionization state. The flux ratios of close lines, which depend
only slightly on interstellar extinction, can be used to
investigate various emission-line objects. These ratios allow
regions of gas ionized by the radiation of OB stars be confidently
distinguished from those where gas excitation is due mostly to
other ionization sources (radiation of the active galactic
nucleus, shocks). At the same time, these diagrams do not always
allow one to separate the contribution of shock ionization from
that of the radiation of old stars of the asymptotic giant branch
or LINER-type galactic nuclei.

\citet{Hong2013} used optical narrow-band photometry acquired with
the Hubble Space Telescope by \citet{Calzetti2004} to analyze the BPT
diagrams of individual regions in NGC\,3077 and demonstrated that
shock ionization contributed to gas ionization in the regions
located closer to the periphery of the galaxy. The above authors
point out that weak manifestations of shocks in ionized gas may be
difficult to discern in the presence of the powerful radiation of
bright  \HII{} regions.

The contribution of shocks to the gas ionization state can be
estimated by supplementing classical BPT diagrams with one
additional parameter---line-of-sight velocity dispersion of
ionized gas ($\sigma$) whose increase is associated with the
increase of turbulent gas velocities behind the shock front.
However, accurate measurements of  $\sigma$ require rather high
spectroscopic resolution $R>5\,000$--$6\,000$. That is why in most
of the surveys such approach is used to study objects with
sufficiently high line-of-sight velocity dispersions (AGNs or
starburst galaxies, where $\sigma>100$--$200\kms$), while being
little used for studying dwarf galaxies and extended regions of
diffuse ionized gas (DIG) with low surface brightness.

When studying such objects \citet{Lopez2017,Oparin2018} used the
line ratios determined with classical methods of integral-field
spectroscopy combined with the data of observations made using
Fabry-Perot interferometer (FPI) with sufficiently high
spectroscopic resolution for estimating the velocity dispersion.

We report the results of a comprehensive analysis of the state of
ionized gas in NGC\,3077 based on new observational data:
long-slit and 3D spectroscopy with the Fabry-Perot interferometer on the 6-m telescope of the Special Astrophysical
Observatory of the Russian Academy of Sciences (SAO RAS) and
narrow-band photometry with MaNGaL instrument attached to the
2.5-m telescope of the Caucasian Mountain Observatory of Sternberg
Astronomical Institute of M.V. Lomonosov Moscow State University
(CMO SAI MSU). We describe our observations and their reduction
uin Section~\ref{sec:obs}. Section~\ref{sec:results} describes the
main observational results. In particular, Section
\ref{sec:kinematics} describes an analysis of the kinematics of ionized gas based on observations made with the FPI;
Section~\ref{sec:ls_results} presents the results of our analysis
of the emission spectrum of the galaxy and of the gas metallicity
based on the data of long-slit spectroscopy;
Section~\ref{sec:mangal_res} considers   the gas  ionization state based on the data of narrow-band photometry
performed with MaNGaL. Section~\ref{sec:summary} summarizes our
conclusions.
%Fig 1

\section{OBSERVATIONS AND DATA REDUCTION}
\label{sec:obs}

\begin{table*}
\caption{Log of observations of NGC\,3077}\label{tab:obs}
    \centering
    \begin{tabular}{c|c|c|c|c|c|c}
        \hline
        Mode & Date & T$_\mathrm{exp}$, s & $\theta$, arcsec & $\Delta\lambda$, \AA & $\delta\lambda$, \AA & Scale, $''$/px \\
        \hline
        LS $PA=312$& Oct 11, 2013 & $6\times1200$ & 1.6 &   3650--7250  &   5.2  &0.35 \\
        LS $PA=234$& Feb 21, 2014& $6\times1200$ & 2.8 &    3650--7250  &   5.2  &0.35 \\
        \hline
        FPI-1   & Nov 09, 2013 & $40\times300$ & 2.0 &8.7 in the vicinity of \Ha & 0.4 & 0.71\\
        FPI-2   & Nov 10, 2013 & $40\times300$ & 1.6 &8.7 in the vicinity of \Ha & 0.4 & 0.71\\
        \hline
        MaNGaL  & Apr 10, 2018& $3\times300$ & 1.8 & \Ha & 15 & 0.66 \\
                & Apr 10, 2018& $6\times300$ & 1.8 & \NII\,$\lambda\,6583$ & 15 & 0.66 \\
                & Apr 10, 2018 & $6\times150$ & 1.8 & \Ha\ cont1 & 15 & 0.66 \\
                & Apr 10, 2018& $5\times150$ & 1.8 & \Ha\ cont2 & 15 & 0.66 \\
                & Apr 10, 2018& $4\times300$ & 1.9 & \SII\,$\lambda\,6717$ & 15 & 0.66 \\
                & Apr 10, 2018& $4\times300$ & 1.9 & \SII\,$\lambda\,6731$ & 15 & 0.66 \\
                & Apr 10, 2018 & $3\times300$ & 1.9 & \SII\ cont. & 15 & 0.66 \\
                & Apr 10, 2018 & $7\times300$ & 2.0 & \OIII\,$\lambda\,5007$ & 15 & 0.66 \\
                & Apr 10, 2018& $5\times300$ & 2.0 & \OIII\ cont. & 15 & 0.66 \\
        \hline
    \end{tabular}
\end{table*}

Table~\ref{tab:obs} lists the details of observations: the mode of observations, date, number and duration of exposures, seeing $\theta$, % seeing
spectral range $\Delta\lambda${} and spectral resolution $\delta\lambda$, and
the image scale.

\subsection{3D spectroscopy with the scanning FPI}
\label{sec:ifp} We investigated the kinematics of ionized gas  in
the \Ha\ emission line in the primary focus of the 6-m telescope
of SAO RAS using SCORPIO-2 multimode focal
reducer~\citep{Afanasiev2011} operating in the scanning
Fabry-Perot interferometer (FPI) mode.  The interferometer  IFP751
provided a free spectral interval of $\Delta\lambda = 8.7$\,\AA\
between neighboring interference orders with a scale of 0.22\,\AA\
per channel. The operating spectral range in the vicinity of the
redshifted  \Ha\ was selected with the \#77B narrow-band filter
with a Central wavelength of\mbox{$CWL=6571$\,\AA.} Scanning
consisted of a sequence of 40 interferograms acquired with
different spacings between the interferometer plates and uniformly
filling the wavelength interval covered. The field of view had the
size of $6\farcm1 \times 6\farcm1$ with a scale of
$0\farcs7$/pixel. The result of reduction performed with the
software  described by \citet{MoiseevEgorov2008,Moiseev2015}
had the form of a data cube where each field-of-view pixel
contained a 40-channel spectrum. Observations were made at two
different orientations of the field of view of the instrument and
the resulting data cubes were then coadded to remove parasitic ghosts
\cite[see][]{MoiseevEgorov2008}.

\subsection{Long-slit spectroscopy}
\label{sec:ls}

We used the same  \mbox{SCORPIO-2} instrument attached to the
\mbox{6-m telescope} of SAO RAS in the long-slit spectroscopy mode
(with the slit \mbox{size of $6\farcm1\times1''$)} to acquire
spectra at two position  angles
($PA=234\degr$ and 312$\degr$) passing through the cavity in the
distribution of ionized gas in the H$\alpha$ line in the northern
part of the central region of the galaxy. We  show the location of
the spectrograph  slits in Fig.~\ref{fig:sdss3077}. We used a
VPHG~1200@540 grating, which provided a spectral resolution of
$\delta\lambda \approx 5$\,\AA\, in the wavelength interval
$\Delta\lambda=3650$--$7250$\,\AA.

We reduced observations in accordance with a standard procedure
using an IDL software  developed for reduction of long-slit
data from SCORPIO-2. The principal reduction stages included bias
subtraction, cosmic-ray hit removal, flatfielding, linearization,
and subtraction of the emission spectrum of the night sky.
Wavelength calibration is performed using the \mbox{He-Ne-Ar}
comparison spectrum taken on the same night. To take into account
the spectral response of the instrument and transform the spectra
to absolute energy units we used the spectrophotometric standards
BD+28d4211 and BD+33d2642 whose spectra were acquired on the same
night as those of NGC\,3077 and at a similar zenith angle.

We used ULySS software
suite\footnote{\url{http://ulyss.univ-lyon1.fr}}
\citep{Koleva2009} to model the spectrum of the underlying stellar
population for each position along the slit and subtracted it from
the acquired spectra. This procedure allowed us to analyze the
emission spectrum of ionized gas in the galaxy. To measure
emission-line fluxes, we fitted the line profile observed at each
position along the slit by a single-component Gaussian using our
IDL software based on MPFIT procedure \citep{mpfit}. The flux
errors presented below in the text and plots are the quadratic
sums of errors due to noise and primary reduction and the error of
the Gaussian fit.

All the fluxes that we report in this paper are corrected for
interstellar extinction inside  NGC\,3077. We determined the color
excess $E(B-V)$ for each region based on Balmer decrement
(\Ha/\Hb) and on a comparison with the theoretical flux ratio
\mbox{\Ha/\Hb~$=2.86$} for $T_e=10000$~K. To correct the fluxes
for  reddening we used the extinction curve of \citet{Cardelli1989}
as parametrized in \citet{Fitzpatrick1999}.

\subsection{Narrow-band photometry with MaNGaL}
\label{sec:mangal}

Narrow-band images of the galaxy in the emission lines
\OIII\,$\lambda\,5007$, \Ha, \NII\,$\lambda\,6583$,
\SII\,$\lambda\,6717,6731$ were acquired with the MaNGaL (Mapper
of Narrow Galaxy Lines) photometer with a tunable filter developed
at SAO RAS \citep{mangal} and attached to the Naysmith-2 focus of
\citep{Kornilov2014}. The instrument is an afocal
reducer with a low-resolution (interference order of about 20 at
\Ha{} line) scanning piezoelectric FPI mounted in front of its
camera in the converging beam. The width of the instrumental
contour (i.e., of the narrow-band filter) in the wavelength
interval employed is \mbox{$FWHM=13\pm1$\,\AA.} The central
wavelength ($CWL$) can be set using  CS-100 
controller\footnote{All the FPIs and controllers used for
observations are manufactured by IC Optical Systems Ltd, UK} with
an accuracy of about 0.4\,\AA. A low-noise $1024\times1024$
iKon-M934 CCD was used as a detector. Observations were made in
the instrumental binning mode (\mbox{$2\times2$}) to reduce noise
and readout time.

Unlike the classical FPI operating in a collimated beam,
the adopted optical layout makes it possible to achieve a
relatively large diameter of monochromatic beam \citep[Jacquinot spots, see][]{Jones2002}. In our case the $CWL$ of the FPI transmission
varies by less than \mbox{$\pm0.5FWHM$} throughout the entire
$5\farcm6 \times 5\farcm6$ field of view. Therefore variations
of $CWL$ along the apparent NGC\,3077 image can be neglected. Note
also that according to our acquired velocity field
(Fig.~\ref{fig:comp_maps}), Doppler variations of the
emission-line wavelengths do not exceed 2--3\,\AA, i.e., they are
small compared to the filter width.

During observations we successively accumulated the images in the
process of the tune-up of the filter to the emission line (taking
into account the average velocity of the galaxy and the
heliocentric correction) and the continuum shifted by
30--50\,\AA. Observations performed in such series allow averaging
the contribution from the variations of atmospheric transparency
and seeing. The transmission peaks from neighboring interference
orders were blocked using intermediate-band filters with a width
of about 250\,\AA. Dedicated filters were used for observations
in the \OIII+continuum, \Ha+\NII+continuum, and \SII+continuum
lines. In the case of  \Ha+\NII\ lines the continuum images were
taken in spectral bands on both sides of the line and then
averaged, whereas in the case of the other lines the continuum
images were taken only in one band.

Reduction of the acquired images differed little from standard
reduction of direct images. After bias subtraction and
flatfielding (illumination of the instrument by an incandescent
lamp through an integrating sphere) the frames acquired in one
spectral band were superposed, and coadded with cosmic-ray hits
removed in the process. Superposition was performed using
reference stars. Continuum images multiplied by a close-to-unity
coefficient determined so that the resulting flux from stellar
images would be equal to zero were then subtracted from the images
taken in emission lines. Continuum was observed at very close
wavelengths and therefore the quality of its subtraction was
appreciably higher than in the case of standard observations with
50--200\,\AA{} wide filters.

We performed absolute flux calibration of the images by observing
spectrophotometric standards using the technique described
in~\citet{mangal}.

The \mbox{\Ha\ emission-line image (Fig.~\ref{fig:sdss3077})} shows
practically all weak emission features seen in the deepest
published images taken in this line with the 6-m
telescope \citep{Karachentsev2007}, but with clearly better
subtraction of foreground stars and stellar population of the
galaxy.  The  \OIII{} image shows numerous compact objects, which
are almost absent in \Ha. We consider them to be candidate
planetary nebulae and supernova remnants. We will investigate them in a
separate paper.

We used Astrometry.net\footnote{http://astrometry.net}
\citep{Lang2010} service to perform astrometric calibration of
MaNGaL images and FPI data cube. This calibration allowed us to
accurately compare the data of long-slit and field
spectroscopy and tunable filter photometry. We then corrected the
map of flux ratios in close \NIIHa{} emission lines based on
MaNGaL data for the effect of low spectral resolution in
accordance with the equation reported in \citet{mangal}.

\subsection{Account of extinction and  \Hb-line brightness map}

For constructing the diagnostic diagrams separating the
contributions from different ionization sources it is important to
measure the  \OIIIHb\ emission-line ratio. We did not perform \Hb{}
line observations with MaNGaL because the observed flux in this
line is several times weaker than the \Ha{} flux and such
observations would require too long exposures. Instead of this we
tried to compute the \Hb{} brightness distribution based on an \Ha\
image and an extinction estimate based on SDSS DR9 photometry data
using two different methods.

Method~1 interprets the decrease of the $g$-band brightness $I$ at
the given point relative to the mean brightness $I_0$ at the
corresponding radius as dust extinction: $A_g=-2.5\lg I /I_0$. We
averaged the $I_0(r)$ profile over ellipses corresponding to the
orientation of outer isophotes ($PA=50\degr$, $a/b=0.85$). We
masked the foreground stars, star-forming regions, and the most
conspicuous dust lanes in the initial image. Because of the
asymmetry in dust extinction we performed averaging separately on
two sides from the major axis and then superposed the two halves
to obtain the final $A_g$ map. We interpreted the brightness
excess relative to  $I_0$ as the absence of extinction.

In method~2 the extinction estimate is based on the  $(g-r)$
color-index map, which is used to construct the average radial
profile of the \mbox{$(g-r)_0$} color index  of the stellar
population obtained  by averaging over the same elliptical rings
as those used in method~1. We similarly masked the peculiar
regions. We determined dust extinction as \mbox{$A_g=3.1 E(g-r)$},
where \mbox{$E(g-r)=(g-r)-(g-r)_0$.}

Both methods may yield biased result in extinction estimates. This
is primarily due to the fact that dust in NGC\,3077 has a complex
and nonuniform distribution compared to stars, it shows individual
filaments intricately associated with  \HII\ regions and diffuse
gas. We therefore compared the maps of ``photometric'' extinction
$A_{g,{\rm phot}}$ obtained by our own  method with spectroscopic
$A_{g,{\rm sp}}$ estimates along two available spectroscopic
cross-cuts computed by the standard Balmer-decrement method as
described above in Section~\ref{sec:ls}. Our comparison showed
that a correction factor $k=A_{g,{\rm sp}}-A_{g,{\rm phot}}$ has
to be introduced. The factor  $k$ so determined also takes in to
account the foreground line-of-sight extinction in the Milky Way.
However, method~1 better describes the variations of extinction
observed along the spectrograph slit and the correction factor is
smaller. We therefore computed the  \Hb\ brightness map using
Method~1.

\section{RESULTS OF OBSERVATIONS}
\label{sec:results}
\subsection{Analysis of the kinematics in the  \Ha{} line}
\label{sec:kinematics}

%Fig 2
\begin{figure*}

\includegraphics[width=\linewidth]{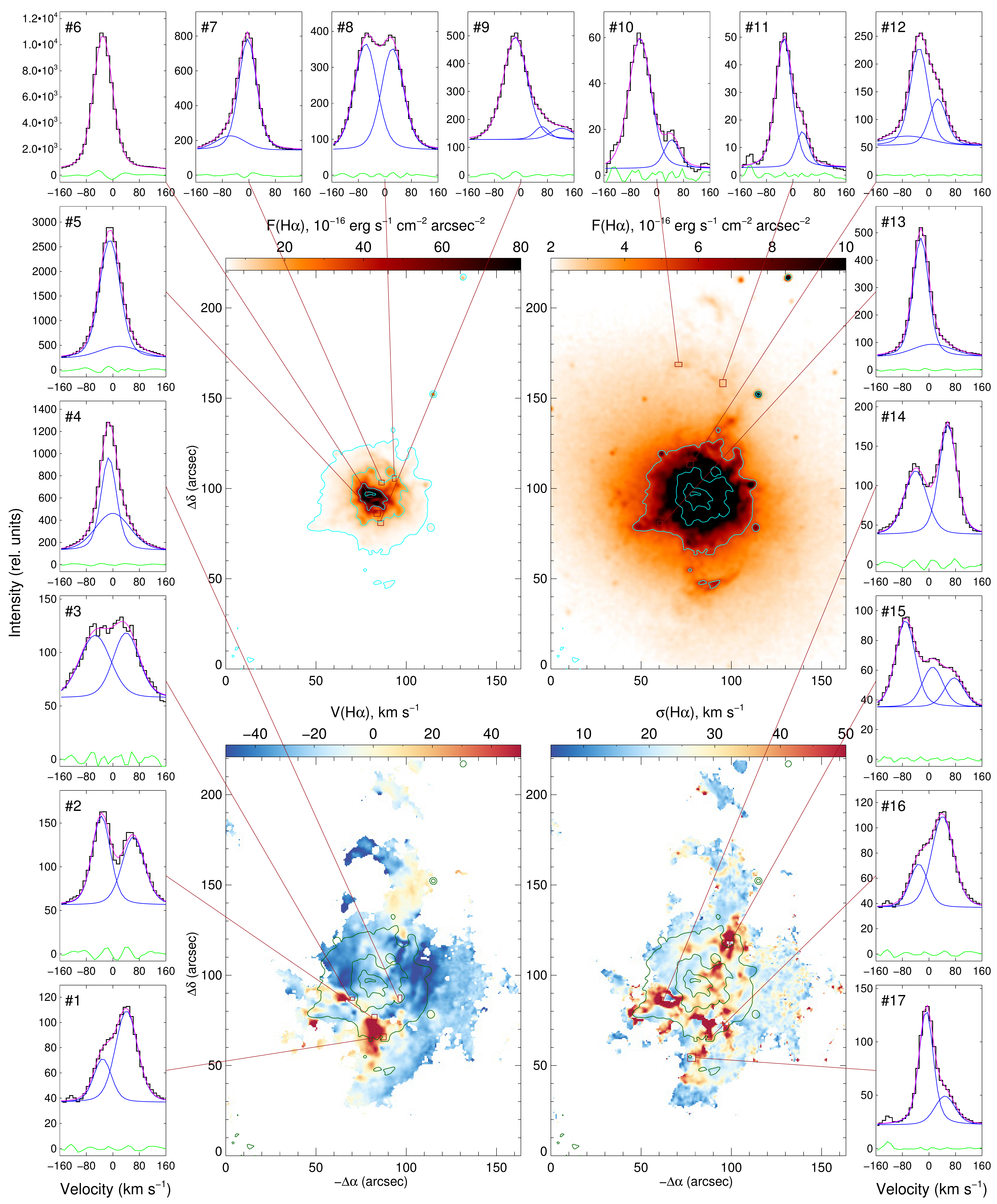}
\caption{Example of observed  H$\alpha$ profiles and the result of
their decomposition into 1--3  Voigt functions (the corresponding
profiles are shown in blue, whereas the green plots show the
residual obtained by subtracting these profiles from the observed
spectrum). The profiles are obtained by integrating over the
rectangular areas indicated in the images. The upper row of images
demonstrates the  H$\alpha$ flux distribution for different
intensity levels. The lower row shows the fields of line-of-sight
velocities (left) and velocity dispersion (right) obtained by
fitting the observed profiles by a single-component Voigt
function. The isophotes correspond to the  H$\alpha$
brightness (0.2, 0.5, 1.6, 4.9, 14.6)$\times
10^{-16}$
erg~s$^{-1}$\,cm$^{-2}$\,arcsec$^{-2}$}\label{fig:Ha_profs}
\end{figure*}

%Fig 3
\begin{figure*}
\includegraphics[width=\linewidth]{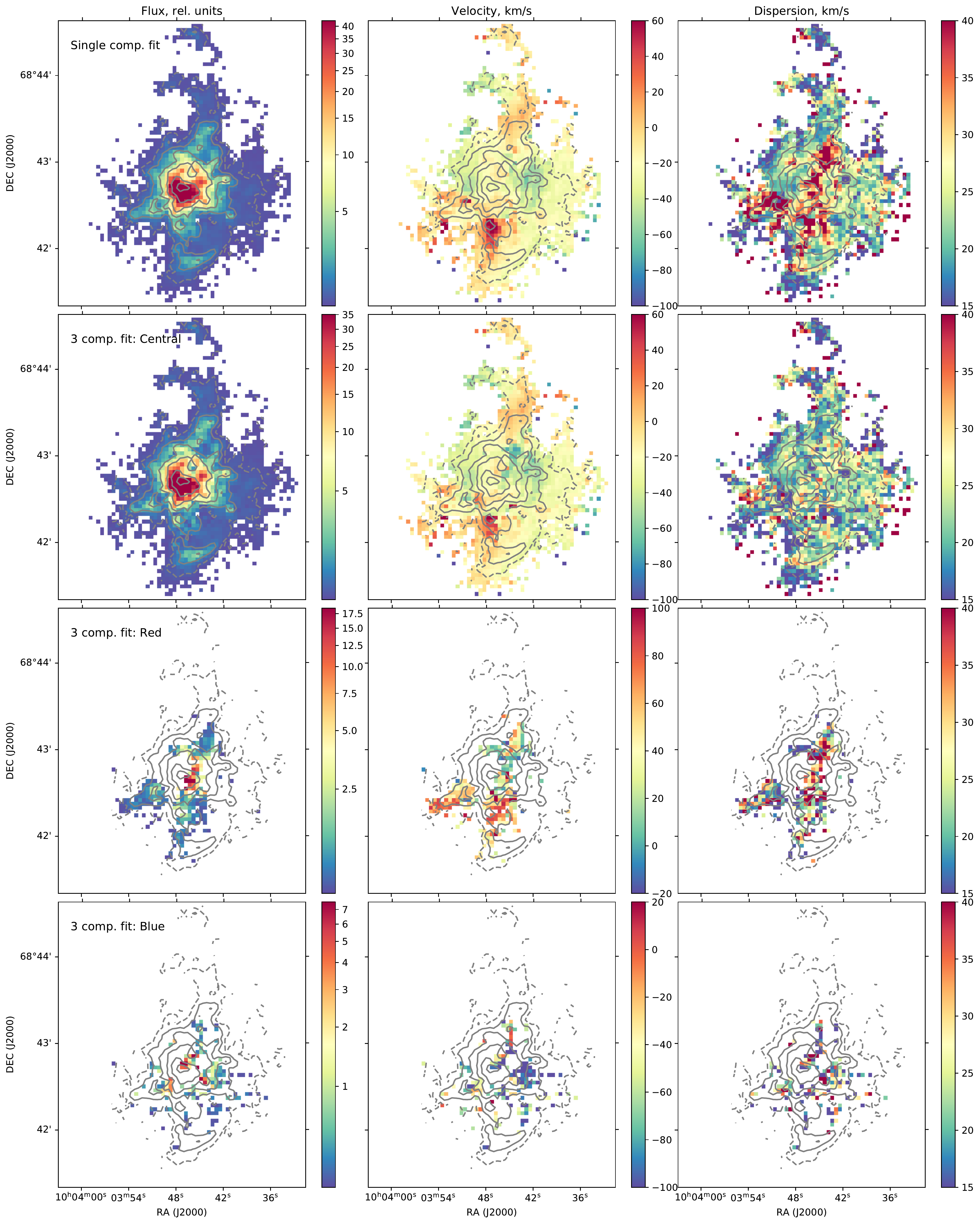}
    \caption{Results of the decomposition of the \Ha{} FPI data
    cube for NGC\,3077 into one and three kinematically decoupled components.
    Left to right---\Ha\ flux, line-of-sight velocity, and velocity dispersion maps.
    The top panels are obtained by decomposing via a single-component Voigt profile,
    whereas the three bottom panels contain information about the central and red- and
    blueshifted components in the case of the three-components decomposition.
    }\label{fig:comp_maps}
\end{figure*}

Emission-line profiles of ionized gas observed with FPI usually
fit quite well the Voigt
function~\citep{MoiseevEgorov2008,Moiseev2015}. The result of 
a fit of individual spectra in the data cube were used to
construct galaxy images in the emission line and in the continuum,
as well as the maps of radial velocity and velocity
dispersion~$\sigma$ free from instrumental broadening. The bottom
panels of Fig.~\ref{fig:Ha_profs} show the maps of the
distribution of radial velocities and velocity dispersion $\sigma$
obtained by fitting a single-component Voigt profile to the data
cube (the top panels show the distribution of \Ha\, surface
brightness according to MaNGaL  observations in the  same scale).

Although the radial-velocity map of the galaxy  shows a certain
($\approx$\,20\,$\kms$) difference in the West to East direction, it
does not exhibit a conspicuous rotational pattern, no rotation
axis can  found, and the circular motion cannot be properly fitted
by a well-defined model. This is consistent with the results  of
\citet{GHASPVI}. The velocity dispersion estimates that we
inferred from \Ha{} data mostly do not exceed
\mbox{20--30~$\kms$,} amounting to  \mbox{35--60~$\kms$} in some
contact regions. In particular, some of these high-dispersion
regions coincide with compact X-ray sources from \citet{Ott2005}.

An analysis of the \Ha{} data cube of  NGC\,3077 revealed numerous
regions exhibiting a complex structure of the emission-line
profile: asymmetry, splitting into two or more components (see
examples of the profile decomposition in Fig.~\ref{fig:Ha_profs}).
The very center of the galaxy exhibits a conspicuous
single-component \Ha{} profile. At the same time, complex
multicomponent profiles are found practically throughout the
entire galaxy outside the brightest  H\,II region. In particular,
in all regions with high velocity dispersion inferred from a
single-component fit \Ha{} profiles are observed that clearly split
into two or three components (see, e.g., profiles \mbox{\#14--17}
in Fig.~\ref{fig:Ha_profs}). A similar pattern is also observed in
regions with the largest radial velocities (profiles \#1--4).

To assess the behavior of gas motions in the galaxy, we tried to
decompose the \Ha{} profiles into kinematically decoupled
components throughout the entire data cube and follow the space
distribution of each component. To this end, we fitted the \Ha\
profile in  each spatial element of the data cube (prebinned with
a 4-pixel bin size to increase the signal-to-noise ratio) by
\mbox{1--3} components. We performed the procedure iteratively
starting with the smallest number of components and adding the
next component in the case if the residual from the subtraction of
the model exceeds the noise level by a factor of three. We
adjusted the number of components manually in some regions based
on the results of a visual inspection of the results of automatic
analysis. We then identified the  ``central'', ``blue'', and
``red'' components (in accordance with their radial velocities)
and analyzed the space distribution of their properties. In the
case of a two-component profile we considered the brightest peak
to be the  ``central'' component. The results are presented  in Fig.~\ref{fig:comp_maps}, the top panel shows the
previous result of fitting by a single component for comparison.

The maps of kinematically distinct components reveal that
practically all high-dispersion regions in Fig.~\ref{fig:Ha_profs}
and in the top panel in Fig.~\ref{fig:comp_maps} show a (mostly)
redshifted or a blueshifted component. The velocity
dispersion of the central component practically coincides
with that of unperturbed regions of the galaxy. Note also that the
locations of the red- and blueshifted components coincide with the
dust lanes seen in SDSS images and also with the superbubbles found
based on the data of echelle spectra~\citep{Martin_1998} .

The complex structure of the line profile makes it
difficult to uniquely separate the components and construct their
 spatial distribution preventing bona fide reconstruction
of the 3D structure of the ionized gas. However, we do not observe the
line-of-sight velocity distribution pattern characteristic of
expanding supershells, where the velocity difference between the
components is maximal at the center of the bubble and decreases
toward the edges. At the same time, we cannot rule out the
existence of compact shell-like structures, like, e.g., those
described in~\citet{Egorov_IC2574, Egorov2018}, or the existence
of expanding superbubbles in significantly nonuniform
medium~\citep{Lozinskaya2003,Egorov2010}.

Note also that some profiles shown in Fig.~\ref{fig:comp_maps}
(\#5, 7, 12, 13), exhibit a broad component with low brightness, which
can actually be identified in a substantial part of the galaxy,
but its contribution was not taken into account in
Fig.~\ref{fig:comp_maps}. Hence we may be dealing with yet another
low-brightness component with $\sigma>$100--150~$\kms$. The
velocity dispersion estimate mentioned above was obtained assuming
that $FWHM$ is comparable to or exceeds the working range of the
FPI and therefore cannot be determined. Note that
\citet{Bresolin2020} recently demonstrated conclusively the
presence of broad components in several giant star-forming regions
in the  M\,101 galaxy. The widths of these components measured by
the zero intensity level ($FWZI$)
could be as large as \mbox{750--1300~$\kms$,} implying a velocity
dispersion \mbox{$\sigma = $100--200$\kms$} (assuming
that $FWHM \approx FWZI/3$). The presence of such a broad
component in regions of ongoing star formation can be associated
with the effect that winds from massive stars have on the
surrounding molecular clouds.

\subsection{Analysis of long-slit spectra}
\label{sec:ls_results}
%Fig 4
\begin{figure*}
\centerline{
\includegraphics[width=0.495\linewidth]{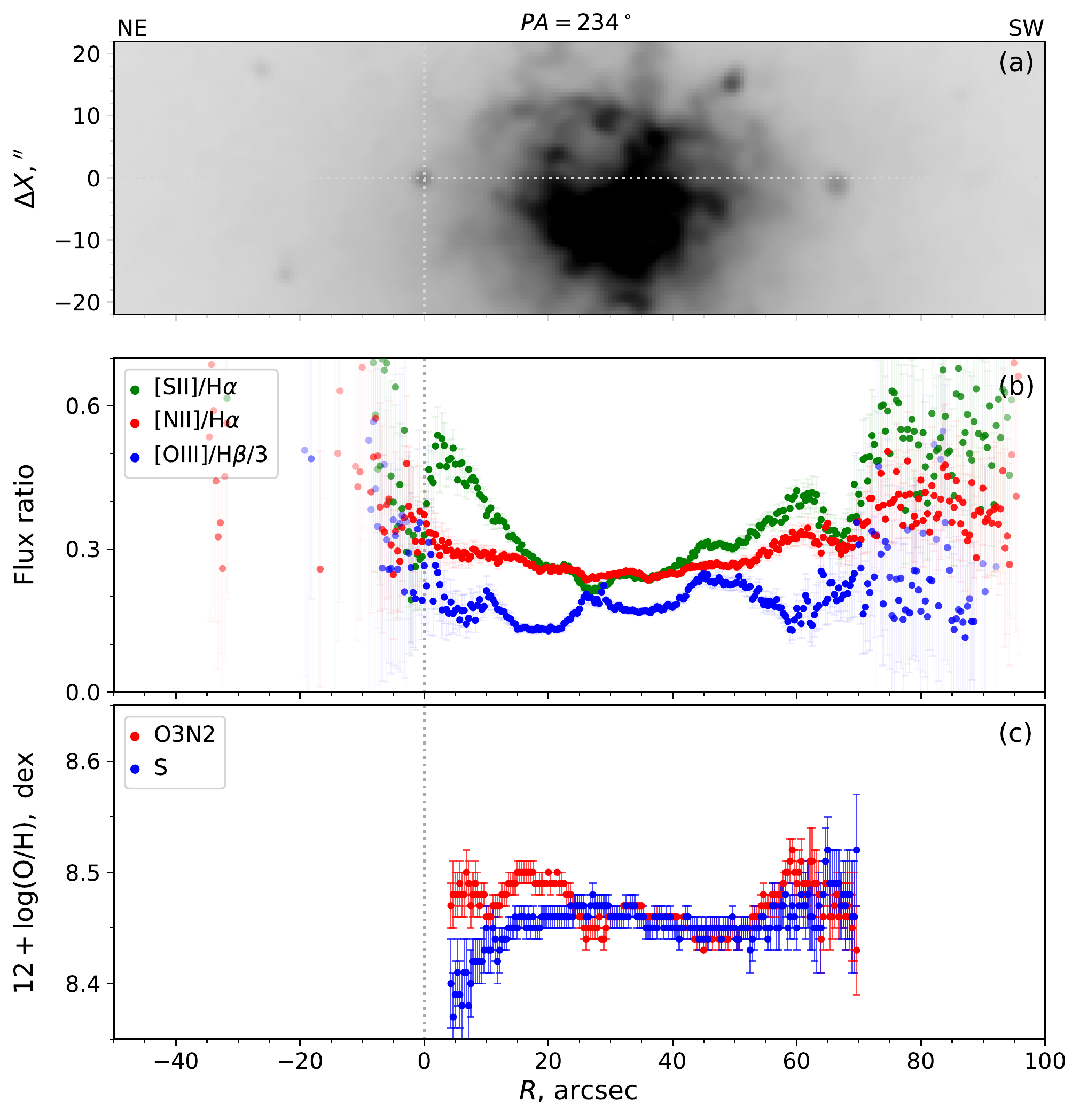}
\includegraphics[width=0.495\linewidth]{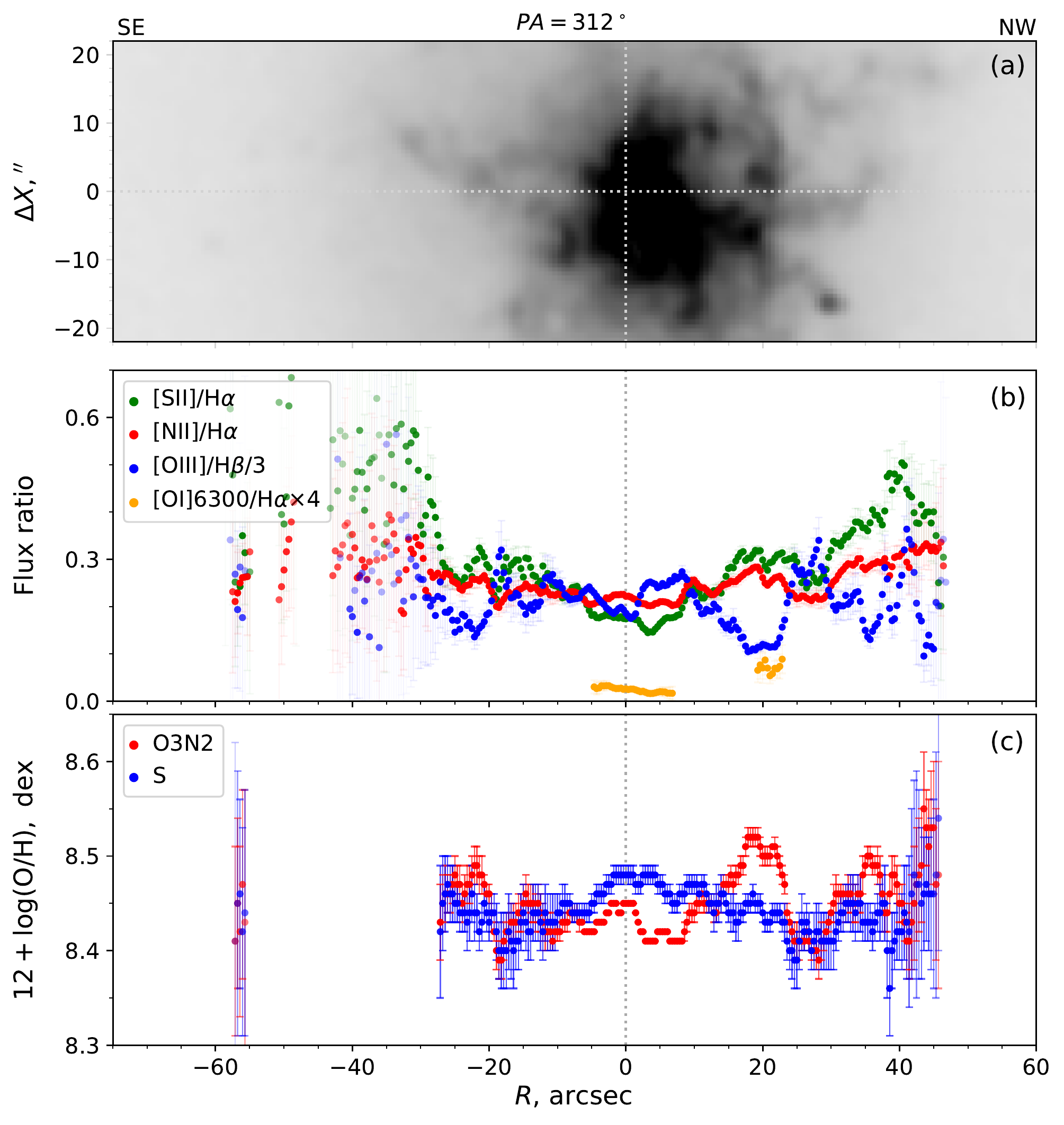}
}
\caption{
Results of an analysis of the emission spectrum acquired
with the spectrograph slit at $PA=234\degr$ (left) and
$PA=312\degr$ (right). The panels---top to bottom---show: (a)
H$\alpha$ map of the galaxy obtained with MaNGaL (the horizontal
line corresponds to the alignment of the spectrograph slit); (b)
the distribution of line flux ratios; (c) the distribution of the
relative oxygen abundance $\mathrm{12+\log(O/H)}$ estimated using
two empirical methods.
}
\label{fig:ls_params}
\end{figure*}

Fig.~\ref{fig:ls_params} shows the results of the analysis of the
flux ratio ratios in some emission lines and the metallicity
distribution along the spectrograph slit whose location is
indicated by the horizontal line in the top  panel.

Until recently, the metallicity of the interstellar medium in
NGC\,3077 was studied only by \citet{Storchi-Bergmann1994} based
on the optical spectrum of the central part of the galaxy in a
large aperture. They estimated the oxygen abundance,
which serves as an indicator of the gas metallicity in the
interstellar medium, to be  \mbox{$\mathrm{12+\log(O/H)} = 8.64$.}
Such an estimate is indicative of the solar abundance of heavy
elements, which is not typical of dwarf galaxies of such
luminosity \citep[the galaxy deviates from the ``luminosity--metallicity'' relation][]{Pilyugin2004}).
\citet{Calzetti2004} used the same spectrum to infer an even
higher abundance of $\mathrm{12+\log(O/H)} = 8.9$. Such a
significant difference is primarily due to the fact that the
authors of the two studies used different families of strong-line
methods (calibrated by  \HII\ regions with bona fide estimates of
electron temperature $T_e$ and from photoionization models), which
can result in discrepancies of up to 0.6~dex \citep[see, e.g.,
][]{Kewley2008}. Our long-slit spectroscopy data allow us not only
to refine the gas metallicity in NGC\,3077, but also investigate
its variation along the slits.

We used two empirical methods to estimate the relative oxygen
abundance $\mathrm{12+\log(O/H)}$: O3N2 \citep{Marino2013} and
S \citep{Pilyugin_S}, which are based on the bright emission-line
ratios \OIII/\Hb, \NII/\Ha\, and \SII/\Ha\, (only in the case of
the S method). Both methods are calibrated by  \HII\ regions with
available bona fide measurements of electron temperature $T_e$ and
metallicity made using the $T_e$ method. We excluded from our
analysis the pixels that do not fall
within the  H\,II regions in BPT diagrams (see below) or having
equivalent widths $EW(H\alpha)< 6$\,\AA. It follows from
Fig.~\ref{fig:ls_params} that both methods yield similar values
with no signs of metallicity gradient, but with small variations,
which can be due to the variation of the ionization parameter
caused by the contribution of DIG or just by the error of the
methods, which amounts to about 0.10--0.15~dex.

The estimates determined by applying the two methods to the
integrated spectrum of the central part ($-5$ to $+7$~arcsec from
the slit center along the slit direction for \mbox{$PA=312\degr$})
are equal to $\mathrm{12+\log(O/H)_{O3N2}}=8.43\pm0.02$ and
$\mathrm{12+\log(O/H)_{S}}=8.47\pm0.03$ (the quoted errors do not
include the errors of the methods). In this region the
signal-to-noise ratio in the blue part of the spectrum was
sufficient for including the [OII]\,$\lambda\,3727$/\Hb{} ratio into
the analysis. We added this ratio to our metallicity analysis and
used the R--method from \citet{Pilyugin_S} to obtain the estimate
$\mathrm{12+\log(O/H)_{R}}=8.49 \pm 0.05$, which is consistent
with the  results mentioned above. Thus unlike what was believed
before the gas metallicity in the galaxy is, according to our
measurements, significantly lower than the solar value and is
equal to $Z=0.6Z_\odot$. We found no significant variations of the
oxygen abundance along the slit. However, as we point out in
Section~\ref{sec:mangal_res} below, more enriched gas associated
with the tidal flow from M\,81, can be observed at the periphery
of the galaxy. Our estimate of the gas metallicity agrees well
with the  ``luminosity--metallicity'' relation
from~\citet{Pilyugin2004} for the adopted NGC\,3077 absolute
magnitude of $M_B=-17.62 $.

\begin{figure*}
\includegraphics[width=\linewidth]{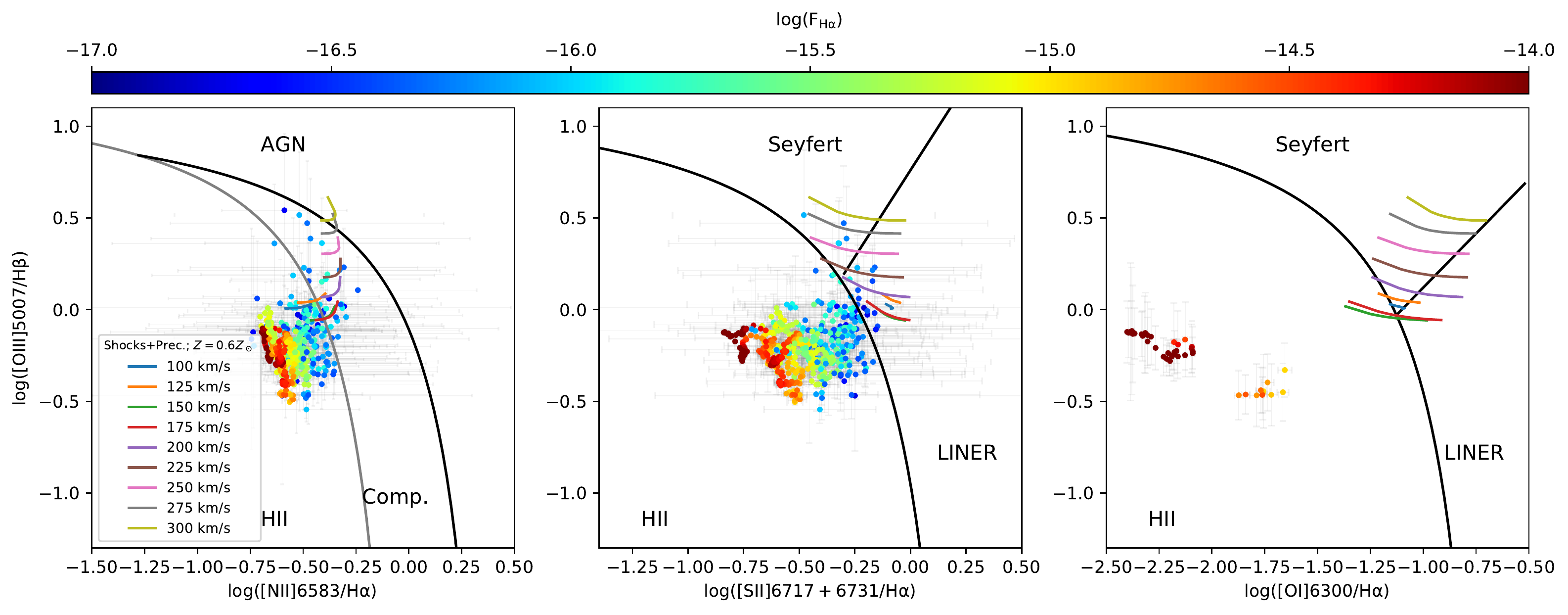}
\includegraphics[width=\linewidth]{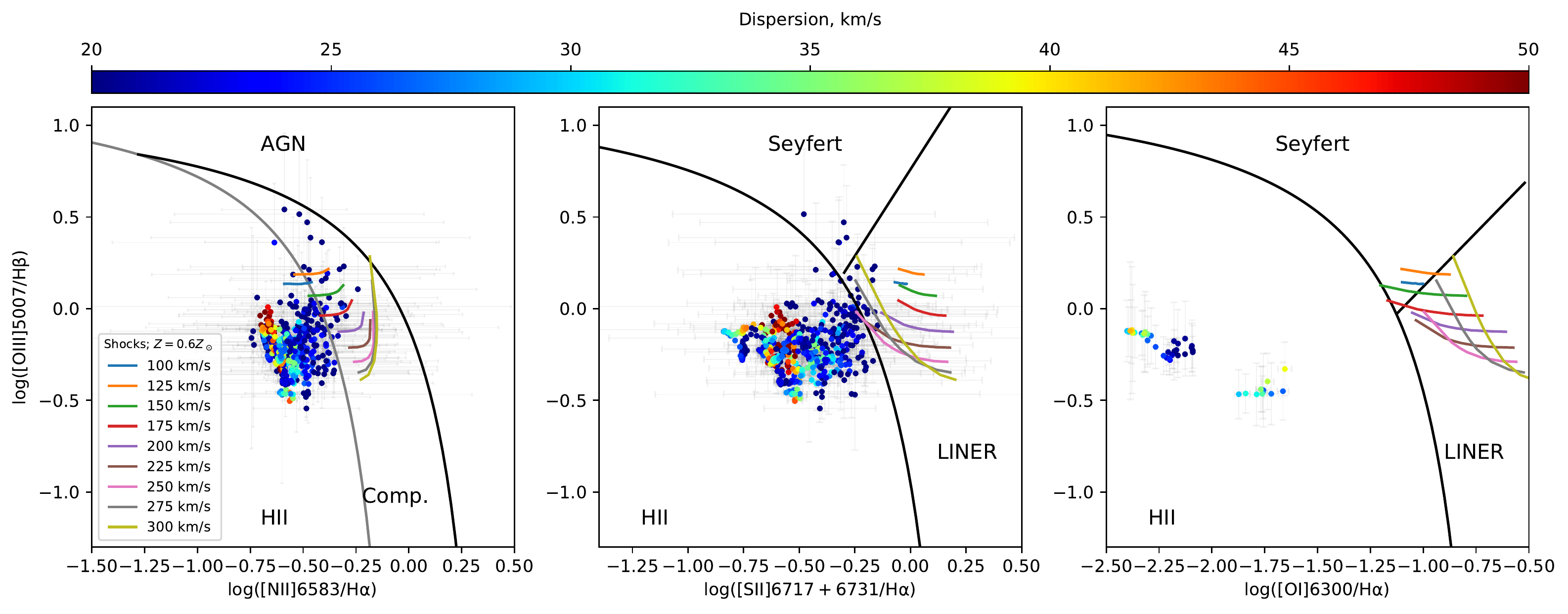}
\caption{BPT diagrams based on spectroscopic observations. Color bar corresponds to   the logarithm of the H$\alpha$ surface brightness
(the top panel) and line-of-sight velocity dispersion (the bottom
panel). The black and gray curves that separate star-forming
regions, composite ionization mechanism, and LINERS are adopted
from~\citet{Kewley2001} and \citet{Kauffmann2003}. The diagonal
line that separates the Seyfert and LINERs regions is adopted
from~\citet{Kewley2006}. The color lines correspond to shock
models in accordance with~\citet{Allen2008} for different
velocities for the metallicity of $Z=0.6Z_\odot$. The top panels
show the models with the precursor and the bottom panels show the
models with shocks exclusively.}\label{fig:ls_bpt}
\end{figure*}

Fig.~\ref{fig:ls_bpt} shows the diagnostic diagrams \OIII/H$\beta$
$vs$ \NII/H$\alpha$, \SII/H$\alpha$, and [OI]/H$\alpha$ based on
the results of spectroscopic observations. The top panels show
color-coded H$\alpha$ surface brightness; the four lines indicate
the position of the curve from \citet{Kewley2001} that separates
the domains where emission can be explained by photoionization
from the domains with another exitation mechanism (AGN, shocks).
The gray line in the left-hand panel from \citet{Kauffmann2003}
separates  \HII{} regions from regions with a composite ionization
mechanism. The straight line in the \OIII/H$\beta$---\SII/H$\alpha$ diagram from~\citet{Kewley2006} separates the
Seyfert and LINER domains. Also shown in the figure are the shock
models adopted from \citet{Allen2008}  for the case of the
metallicity of $Z=0.6Z_\odot$ corresponding to our estimate for
NGC\,3077. Various color curves correspond to models with
different shock velocities. Note that the top panels show the
models with the precursor contribution, whereas the bottom panels
show the models including the contribution from shocks
exclusively. It follows from the diagrams that the entire bright
emission is associated with photoionization by massive stars in
star-forming regions, whereas the domain of composite excitation
is populated only by the points corresponding to low surface
brightness DIG. As is evident from a comparison of diagnostic
diagrams with the models of shocks, their effect can explain the
ionization of low surface brightness regions (the discrepancy
between shock models and observations in the case of
[OI]/H$\alpha$ is due to the lack of [OI] line data for DIG
regions). Note that the shock models without the photoionziation
precursor are worse to explaining observational data. This fact
leads us to conclude that the leakage of ionizing photons from
star-forming regions play an essential role in the ionization of
DIG in the regions located on the slit. In this case shocks did
not play an important role in the ionization of observed bright
regions of ionized gas including extended filaments and superbubbles.

The bottom panels in Fig.~\ref{fig:ls_bpt} show the BPT-$\sigma$
diagrams with color coded line-of-sight velocity dispersion
(determined from FPI spectra as a result of a single-component
profile decomposition) measured along the spectrograph slit. As is
evident from the data shown in the diagrams, in the case of
NGC\,3077 regions of high velocity dispersion are associated with
star-forming regions and not with shocks. However, the
\SII/H$\alpha$ diagram shows a small shift of high-dispersion
regions toward higher line ratios.
%Fig 5
%Fig 6
\begin{figure*}[!]
\includegraphics[width=\linewidth]{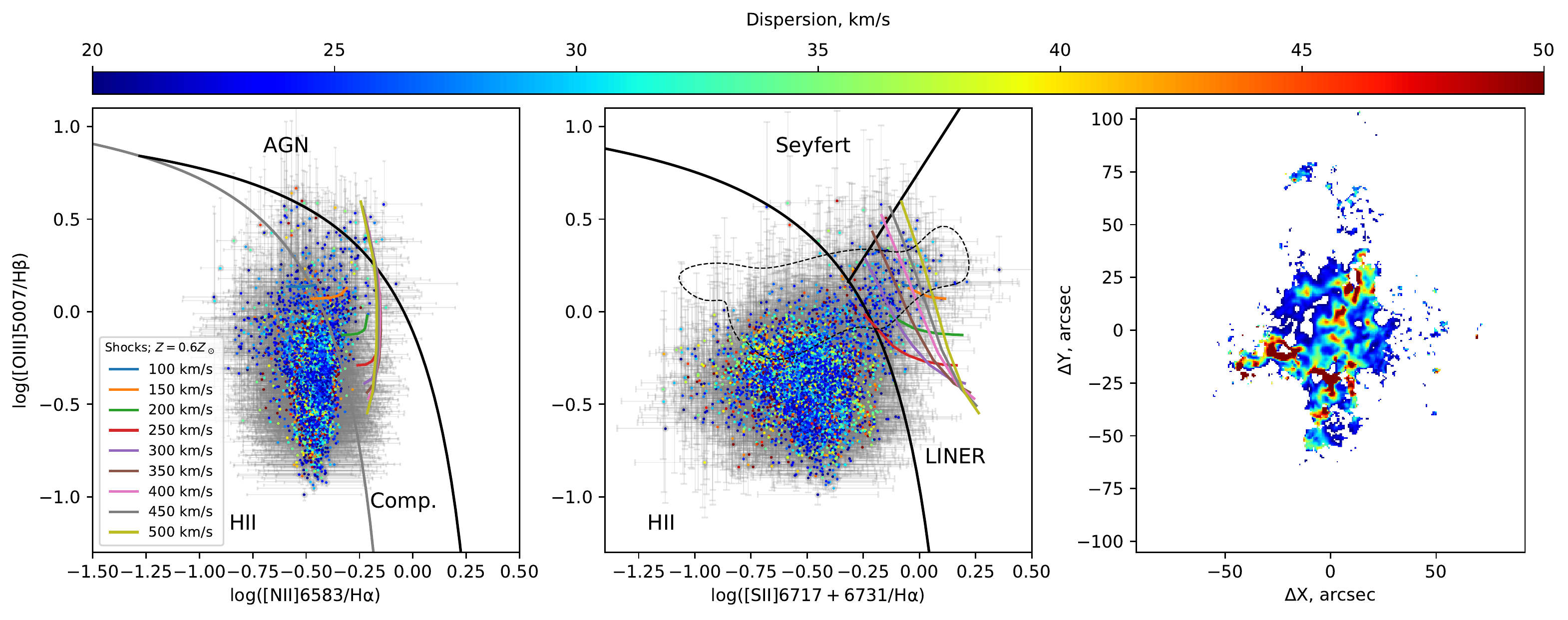}
\includegraphics[width=\linewidth]{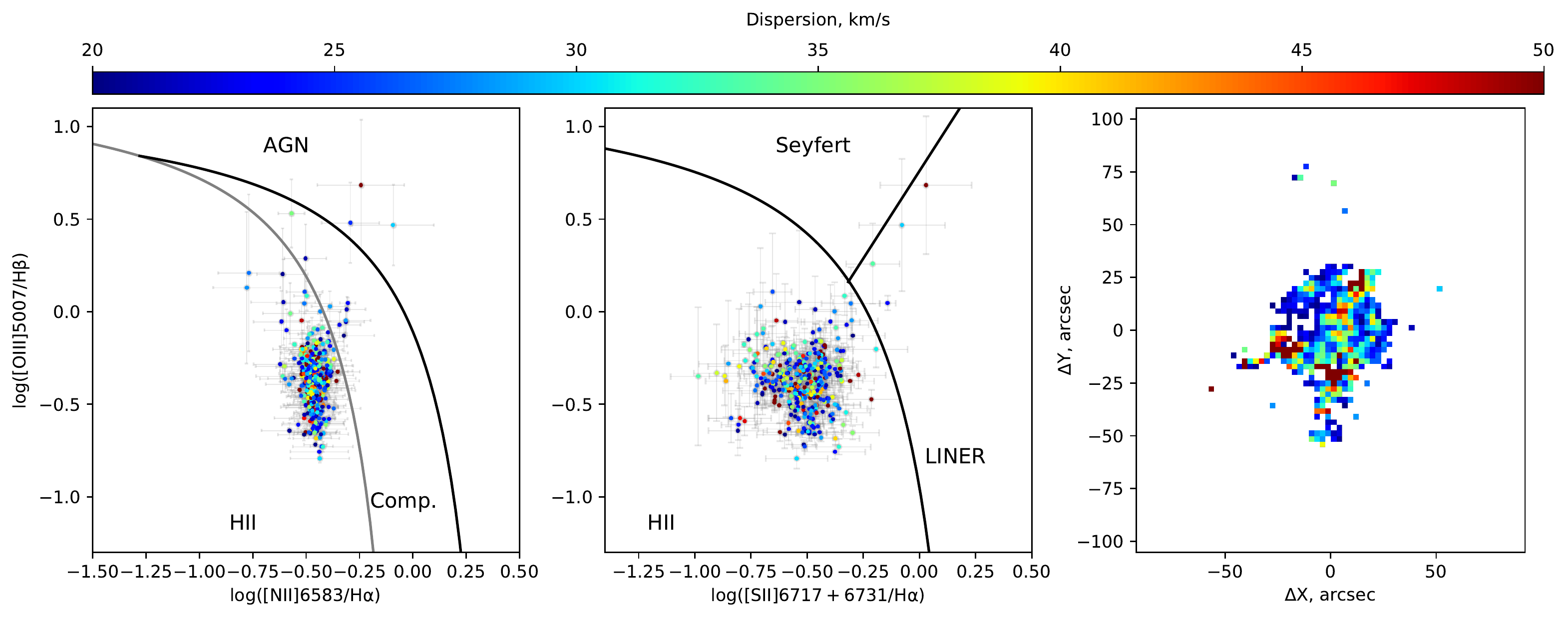}
\includegraphics[width=\linewidth]{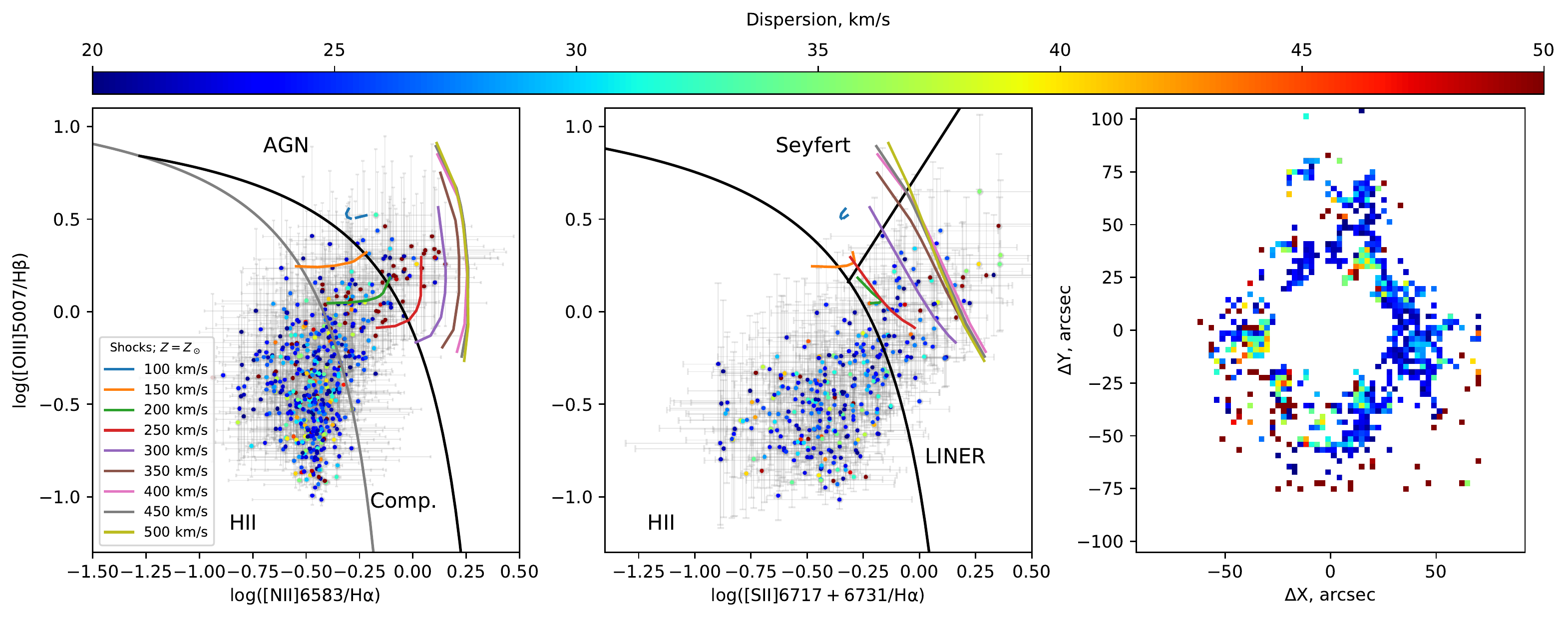}
\caption{The BPT diagrams based on the results of observations
with a tunable filter. The top panels show the result obtained for
the entire galaxy with the original pixel size. The middle and
bottom panels show the diagrams for the inner and outer regions
with binning=4. The colors are used to code the \Ha\ velocity
dispersion, the right-hand panels show the corresponding velocity
dispersion maps for each set of regions. The curves separating the
domains of the diagram with different gas ionization mechanisms
are similar to those shown in Fig.~\ref{fig:ls_bpt}. The color
curves correspond to shock models in accordance
with~\citet{Allen2008} for different velocities and metallicities
$Z=0.6Z_\odot$ (the top panels) and $Z=Z_\odot$ (the bottom
panels). The dashed line in the top middle panel outlines the
domain of observed line flux ratios according
to~\citet{Hong2013}.}\label{fig:mangal_bpt}
\end{figure*}

\subsection{BPT diagrams based on narrow-band photometry}
\label{sec:mangal_res}

The BPT diagrams based on the data of long-slit spectroscopy
cannot describe the state of ionized gas throughout the
entire galaxy and therefore we reconstruct the emission-line
ratios from the results of photometry with a tunable filter.

Fig.~\ref{fig:mangal_bpt} shows the diagnostic BPT diagrams based
on the \OIII/\Hb, \NII/\Ha{}, and \SII/\Ha{} line ratios obtained
via mapping with  MaNGaL. The color coding corresponds to
line-of-sight velocity dispersion estimates inferred from FPI
data. The figure shows three sets of diagrams for different
regions of the galaxy (see below). The right-hand panels show the
maps of line-of-sight velocity dispersion included into each set.
We adopted the parametrization of the curves that separate regions
with different type of ionization  (photoionization regions---H\,II, objects with composite ionization type---Comp., active
Seyfert galaxies---Seyfert, and LINER type objects)
from~\citet{Kewley2006}.  The color lines, like in
Fig.~\ref{fig:ls_bpt}, correspond to shock models for different
velocities adopted from~\citep{Allen2008}, but the bottom panel
shows the values for solar metallicity (in the top panel, for
\mbox{$Z=0.6Z_\odot$).} The top panel in
Fig.~\ref{fig:mangal_bpt} shows the diagrams obtained for the
entire galaxy based on the data with original pixel size. The
absolute majority of the data points in the diagrams lies in  the
domains characteristic for photoionization by the radiation of OB
stars and show no appreciable contribution from shock ionization.
Only a small fraction of the data points is located in regions of
composite ionization. This result agrees with the results of our
long-slit observations, however, compared to the diagrams reported
by \citet{Calzetti2004, Hong2013} our data points are shifted
downward. In our opinion, this is due to the specificities of the
technique of the photometry in narrow- and intermediate-band
filters used on the Hubble Space Telescope. In particular, in HST
observations the contribution of continuum radiation of the
stellar population was taken into account by subtracting from
\OIII\- and \Hb{} line data the F547M-band frames whose effective
wavelength is rather far from the above lines. The technique of
accounting for the contribution of the continuum based on two
images taken at close wavelengths that we employed in our
observations with a tunable filter allowed us to more accurately
estimate the fluxes in emission lines. Furthermore, our spectra
show that the effect of the absorption line from stellar
population is quite significant in the region of the  \Hb\ line.
This effect is difficult to properly account for in filter frames
and that must explain the overestimated \OIII/\Hb{} ratios
reported in the above papers. Our technique for computing the
distribution of the \Hb{} flux from the  \Ha{} image using the
extinction map is free from this disadvantage as confirmed by the
agreement between the line ratios based on SCORPIO-2 data, which
take into account the model of the stellar population, and the
ratios based on MaNGaL photometry.

The surface brightness in the \OIII$\lambda\,5007$ line in outer
regions of the galaxy is appreciably lower than the surface
brightness in \NII\ and \SII{} lines, which is typical for DIG. Therefore, the outer regions of the galaxy, where
brightness in the oxygen line is lower than the noise level but the
signal is quite appreciable in the nitrogen and sulfur lines, are
excluded from the analysis of BPT diagrams. To overcome this
limitation, we artificially fixed the \OIII\,$\lambda\,5007$ line
flux at the level equal to the standard deviation of noise level
for the regions were the  \NII\ and \SII\ are reliably detected,
and  \OIII{} line is undetected. Thus the fixed  \OIII{} line flux is
equal to its upper boundary. Such a procedure allows us to follow
the variations of the excitation mechanism by analyzing the
variations of the  \NII/\Ha\ and \SII/\Ha\ ratios on the BPT
diagrams including the regions with insufficiently high  $S/N$
ratios in the \OIII\,$\lambda\,5007$ line.
Fig.~\ref{fig:mangal_bpt} shows the BPT diagrams and the map of
velocity dispersion for ``inner'' regions of the galaxy where
\OIII\,$\lambda\,5007$ line is securely detected (the middle
panel), and also for outer regions where $S/N\leq 3$ (the bottom
panel).

As is evident from the middle panel in Fig.~\ref{fig:mangal_bpt},
emission in the central part of the NGC\,3077 can be fully
explained by photoioniziation by OB stars, and this also applies
to regions with high velocity dispersion. On the other hand, the
outer parts of the galaxy (the bottom panels) show a well-defined
trend toward the right-hand part of the BPT diagram, beyond the
region of photoionization by young stars, which lies below the
demarcation line between the Seyfert and LINER domains. Our
adopted \OIII\,$\lambda\,5007$ line flux in outer regions is an
upper estimate and therefore in reality the data points on the BPT
diagram should be located no higher than they are located in
Fig.~\ref{fig:mangal_bpt}.

Note that the observed position in BPT diagrams in the LINER
domain can be explained quite well by shock models
\citep{Allen2008} both in the case of the metallicity close to our
estimate for NGC\,3077 (see models in  the top panel) and for
$Z=Z_\odot$ (see the bottom panel). Adding photoionization
precursor to these models allows to even better explain the flux
distribution in the left-hand panel, but the agreement is
significantly worse for \SII/\Ha, and this fact is indicative of a
significant contribution of shocks to ionization in outer regions.
No correlation with the velocity dispersion of ionized gas is
observed in this case. For outer regions the positions of data
points in the BPT diagram can be better reproduced  in the case of
significantly lower shock velocities if shocks propagate through a
medium with solar content of heavy elements than if we assume the
metallicity at the periphery to be equal to our estimate for the
central part based on long-slit spectroscopy.

An analysis of the BPT diagram for the periphery of NGC\,3077
suggests that we observe there a significant contribution of
shocks propagating through a medium with possibly more metal-rich
gas compared to the central part of the galaxy. Such a pattern can
be explained both by the interaction between the galaxy and a
giant tidal \HI{} structure, and by manifestations of the galactic
wind ejecting metal-enriched to the periphery. Both processes can
be responsible for shocks and, possibly, for higher metallicity of
gas at the periphery. However, in the case of the external
accretion of gas clouds we can hardly expect high-velocity shocks:
according to our analysis of the maps published in
\citet{Walter2002, Walter2008}, the line-of-sight velocity of
\HI, clouds is of about $-5\pm15\kms$, which agrees well with the
systemic velocity of NGC\,3077. The hypothesis that mutual
motions occur in the sky plane is too daring. On the other hand,
in the case of significant mass outflow we could expect
appreciable manifestations of shocks toward the center of the
galaxy. However, BPT diagrams in the central part of the galaxy
show no signs of shocks and the main kinematically distinct
supersonic gas motions are observed along large-scale gas-and-dust
filaments or are associated with local X-ray sources (supernovae)
and not with the large-scale gas outflow from the galaxy. We can
conclude that photoionization by the UV radiation of young stars
in the central part of NGC\,3077 is evidently the main contributor
to gas ionization compared to shocks. Note that such a pattern is
typical for known cases of galactic wind: the \NIIHa{} ratio
corresponds to its value for  \HII{} regions in the disk of the
galaxy, but increases toward the periphery with increasing
distance from central star-forming regions (see, e.g., \citealt{Lopez2017,LopezCoba2019}). Note that the presence of a broad
low-contrast component in the \Ha{} line in central regions of the
galaxy is indicative of the outflow of high-velocity gas from
star-forming regions.

Despite the manifestations of gas outflows mentioned above we see
no appreciable correlation between the line ratio and $\sigma$ observed in dwarf galaxies with ongoing star
formation, primarily in those with galactic wind
\citep{Lopez2017,Oparin2018}. This must be due to the fact that
what we see along the line of sight a mix of several
emission components: gas in star-forming regions, gas
flowing out from the disk, and accretion flows toward the disk. We
observe the multicomponent structure of profiles in FPI data.
However, we so far could not uniquely associate the velocity
dispersion of each component with its individual  diagnostic line ratio,
because the observed line ratios were measured in low-resolution spectra. Perhaps it
will be possible to do it later using additional spectroscopic
data and new algorithms of profile decomposition.

\section{CONCLUSIONS}
\label{sec:summary}

We used observations made with the  \mbox{6-m} telescope of SAO
RAS and \mbox{2.5-m} telescope of Caucasian Mountain Observatory
of Sternberg Astronomical Institute of M.V. Lomonosov Moscow State
University using the methods of long-slit and 3D spectroscopy and
narrow-band photometry in emission lines to investigate ionized
gas in the NGC\,3077 galaxy.

Observations with a scanning FPI allowed us to map the
distributions of line-of-sight velocities and velocity dispersion
of ionized gas with high spatial resolution. We used the results of
long-slit spectroscopy to estimate the relative oxygen abundance
in the interstellar medium of the central region of the galaxy,
$\mathrm{12+\log(O/H)}=8.43$--$8.49$~($Z\approx0.6Z_\odot$), which
is significantly lower than the earlier estimate  $Z=Z_\odot$ and
agrees with the expected value for NGC\,3077 given its luminosity.
We detected no radial metallicity gradient. The diagrams of the
emission-line ratios based on the data of long-slit spectroscopy
and photometry with a tunable filter allowed us to refine the
results earlier published by \citet{Hong2013}. The good agreement
between the BPT diagrams that we obtained using two different
methods shows that the use of photometry with a tunable filter to study
the ionization state can be a good alternative to classical
spectroscopic methods for investigating the sources of gas
ionization in extended objects including objects with low surface
brightness.

Despite the presence in NGC~3077 of some regions with relatively
high line-of-sight velocity dispersion of ionized gas---up
to \mbox{$\sigma\approx60 \kms$}, there is no correlation
between the location of points in the BPT diagram and velocity
dispersion $\sigma$,  like in the case of the
dwarf galaxy  VII~Zw~403 \citep{Oparin2018}. In the central parts of NGC\,3077 no
appreciable contribution of shocks to gas ionization can be seen.
This must be due to the strong photoionization just in
star-forming regions: a weak contribution of shocks to the
excitation of [N\,II] and [S\,II] can possibly not show up in the
wings of the bright  \Ha, line shifting the  [N\,II]/\Ha{} and
[S\,II]/\Ha{} ratios leftward in the BPT diagram. In this case
high velocity dispersion in  \Ha\, can be explained by the
superposition of several components along the line of sight. The
decomposition of the data cube obtained with FPI in the \Ha, line
allowed us to investigate the distribution of at least three
kinematically distinct components in the galaxy. The results
showed in many regions the contribution of the broad ($\sigma>100
\kms$) low surface brightness component, which is probably
associated with the manifestation of winds from massive stars in
star-forming regions.

The inferred picture of the distribution of line-of-sight
velocities suggests that a part of the regions with a conspicuous
multicomponent profile that were earlier found based on the data
of echelle spectra~\citep{Martin_1998} are separate kinematic
components rather than  single expanding superbubbles, and this
conclusion is also confirmed by the fact that some of them
coincide with the dust lanes in optical images. In our opinion, we
are dealing with the infall of cold gas from a tidal structure
onto the disk of the galaxy. This infall triggers star formation
at the center of NGC\,3077.

\begin{acknowledgements}
We are grateful to A.~M.~Tatarnikov and
N.~I.~Shatskii for organizing observations at Caucasian Mountain
Observatory of Sternberg Astronomical Institute; to
V.~L.~Afanasiev and A.~N.~Burenkov for conducting observations
with the 6-m telescope of the Special Astrophysical Observatory of
the Russian Academy of Sciences, and to T.~A.~Lozinskaya for her
assistance with data acquisition and discussion of the work.
This research has made use of the NASA/IPAC Extragalactic Database
(NED), which is operated by the Jet Propulsion Laboratory,
California Institute of Technology, under contract with the
National Aeronautics and Space Administration and public data of
the  SDSS  survey (site {\url{http://www.sdss3.org/}}, which us
funded by the Alfred P. Sloan Foundation, SDSS Participating
Institutions, National Science Foundation, the US Department of
Energy, the National Aeronautics and Space Administration (NASA),
the Japanese Monbukagakusho, the Max Planck Society,
 and the Higher Education Funding Council for England.

\end{acknowledgements}

\section*{FUNDING}
This work was supported by the Russian Science Foundation (project
no. 17--12--01335 ``Ionized gas in galactic disks and beyond the
optical radius'').  
Observations on  the telescopes of SAO RAS are carried out with
the supporft of the Ministry of Science and Higher Education of
the Russian Federation (contract No. 05.619.21.0016, project ID
RFMEFI61919X0016).

%\bibliographystyle{aa}
%\bibliography{N3077}

\end{document}